\documentstyle[12pt,epsfig]{article}
\begin{document}
\renewcommand{\thefootnote}{\fnsymbol{footnote}}
\begin{center}
{\large\bf Chaotic behaviour of nonlinear waves and solitons of 
perturbed\\
Korteweg - de Vries equation} \vskip 4mm Konstantin B. Blyuss
\footnote{Corresponding author. E-mail: kblyuss@yahoo.com} \vskip
3mm {\it Department of Theoretical Physics II, Brandenburg Technical
University,}

{\it Postfach 101344, 03013 Cottbus, Germany}
\end{center}

\begin{abstract}
This paper considers properties of nonlinear waves and solitons of
Korteweg-de Vries equation in the presence of external
perturbation. For time-periodic hamiltonian perturbation the width of the stochastic layer is calculated. The conclusions about chaotic behaviour in long-period waves and solitons are inferred. Obtained theoretical results find experimental confirmation in experiments with the propagation of ion-acoustic
waves in plasma.
\end{abstract}

\section{Introduction}

Recently a significant attention is paid to the studying of soliton equations under external perturbations. For KdV equation, in particular, it was shown, that simple sine driven force together with dissipation can lead to a chaotic dynamics, detected by Fourier spectrum, positive Liapunov exponents and phase plane analysis \cite{paper1,paper2}. In this context the reduction of a perturbed KdV equation in the form of stationary travelling waves was considered. The justification of the application of low-dimensional chaos theory to such an infinite-dimensional system as KdV equation is in a numerical investigation, which proves that information dimension, calculated by Kaplan-Yorke formula, and correlation dimension within Grassberger-Procaccia method, are both between 2 and 3 for steady waves \cite{paper3}. On the other hand, perturbed KdV equation can be considered directly whithin numerical approach, or by perturbation technique \cite{paper4}. Besides this series of work it is worth to mention here the cases, when KdV equation is influenced by dissipation and instability, wihtout explicit spatial or temporal dependence \cite{in-proc1}. In such a form this equation describes current-driven ion-acoustic instability in a collision-dominated plasma. For the strongly dissipative case overall evolution demonstrates irregular behaviour, therefore leading to nonstationary and irregular soliton interactions.

In this work we add hamiltonian deterministic perturbation to the KdV equation and study physical consequences of this. The reduction in the form of travelling waves is made in order to turn the system into a second-order ordinary differential equation. The existence of solutions to such an equation was studied in \cite{paper5}, while the construction of periodic orbits can be found in \cite{paper2} for nonresonant and primary resonant cases, and in \cite{paper1} for secondary resonances. We'll, however, perform the analysis of chaotic properties in this situation. It is known, that for any hamiltonian perturbation, which makes the system near-integrable, a stochastic layer around a separatrix appears. This layer is bounded by unbroken KAM-surfaces, and whithin it the system evolves with mixing. For dissipative perturbations, however, these KAM-surfaces are broken and special methods like Melnikov one should be used to determine the conditions for chaos to appear. So, we calculate the width of a stochastic layer, which contains long-period waves and solitons. It is done on the basis of Chirikov criterion for the overlap of resonances, which is widely used for investigation of chaotic properties in hamiltonian systems ( references on original papers in this field can be found in \cite{book1}). Results we obtain are as follows: solitons and nonlinear waves prove to be chaotic in the meaning that in a small distance from the peak of solitons and long-period waves there must be a region of chaotic dynamics, where these waves acquire small irregular deviations. This conclusion is confirmed in experiments with ion-acoustic waves in plasma, where a weak splash before soliton peak was registered \cite{in-proc2}.

The outline of this paper is as follows. In next Section KdV
equation is reduced to ODE and the unperturbed solutions of the latter are
considered. In Section 3 we introduce canonically-conjugated
variables action - angle and obtain general expressions for the
criterion of stochasticity. After that in Section 4 this
criterion is applied to KdV waves directly. Section 5 contains
the conclusions and summary.

\section{Stationary waves}

Let's consider the perturbed KdV equation, taken in the following
form:
\begin{equation}\label{EQ1}
U_t+U\cdot U_x +\beta U_{x x x }=V(U,U_t,U_x,x-vt),
\end{equation}
where $x $ and $t$ denote, respectively, a one-dimensional space
coordinate and time; $U(x ,t)$ is supposed to be differentiable
with respect to $x $ and $t$ sufficient number of times; $V(U,U_t,U_x ,x-vt)$ is a small external perturbation. As far as we restrict ourselves on the case of hamiltonian perturbation, it means that within the class of functions $V$ we'll consider only those, which will not contain derivatives after integration over $(x-vt)\equiv\xi$. It's easy to check that the general form of such perturbations can be represented as:
\begin{equation}
V(U,U_t,U_x,\xi)=f_{1}(U,\xi)\cdot U_t(x,t)+f_{2}(U,\xi)\cdot U_x(x,t)+f_{3}(U,\xi),
\end{equation}
with the condition $\left[ f_2(U,\xi)-c\cdot f_1(U,\xi)\right]_{\xi}=f_3(U,\xi)_U$.

Searching for a solution of equation (\ref{EQ1}) in the
form of a nonlinear stationary wave $U(x ,t)=f(\xi)$, we
obtain the following ODE after one integration over $\xi$ ( prime means differentiation with respect to $\xi $ ):
\begin{equation}\label{EQ2}
\beta f^{^{^{\prime \prime}}}=vf-\frac{f^2}{2}+F(f,f^{^{\prime }},\xi ),
\end{equation}
where $F$ denotes the primitive of the function $V$ after stationary wave substitution. Integration constant is absent here because it can be turned into zero by an appropriate
choice of variables. For convenience let's rewrite equation (\ref{EQ2}) as the following dynamical system:
\begin{equation}\label{EQ3}
\beta \ddot x=vx-\frac{x^2}{2}+F(x,\dot x,t).
\end{equation}
Equation (\ref{EQ3}) without perturbation ( $F(x,\dot x,t) =0$ ) is
analogous to the equation of motion:
\begin{equation}\label{EQ4}
\beta \ddot x=vx-\frac{x^2}{2}
\end{equation}
of a point with mass $\beta$ in a potential field
\begin{equation}\label{EQ5}
U(x)=\frac{x^3}6-\frac{vx^2}2.
\end{equation}
\begin{figure}
\epsfig{width=4cm,file=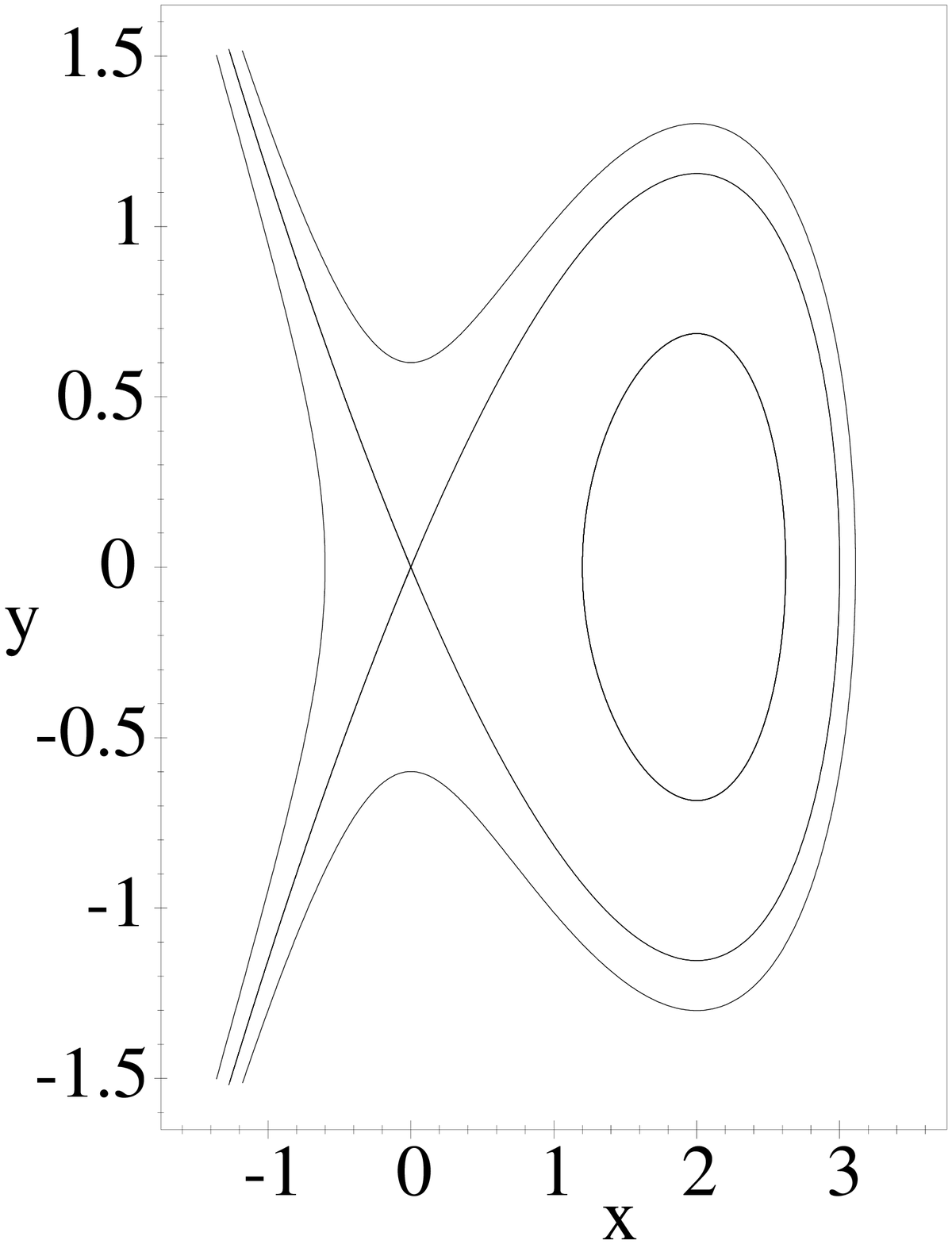}\hspace{2cm}
\epsfig{width=4cm,file=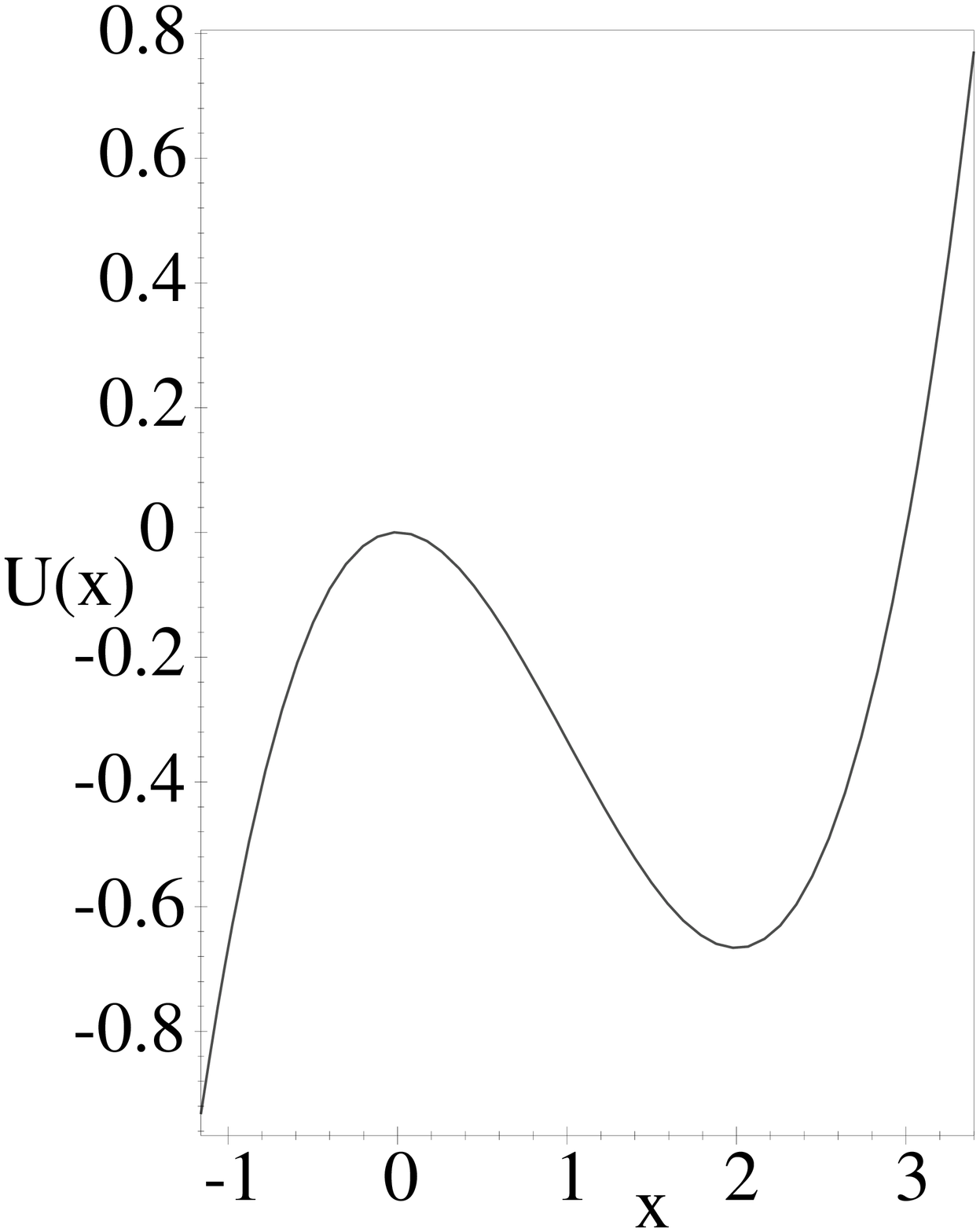}
\caption{(Left) Phase plane of the unperturbed system (\ref{EQ4}). (Right) Plot of the potential energy (\ref{EQ5}). Here $v=1$ and $\beta =1$.}
\end{figure}
Full mechanical energy of such a nonlinear oscillator is equal to
\begin{equation}\label{EQ6}
H=\frac{p^2}{2\beta}-\frac{vx^2}2+\frac{x^3}6,
\end{equation}
where $p=\beta \dot x$ is the momentum of the point.
Phase plane of this oscillator together with the plot of the potential energy is presented on the Fig. 1 .

If $H=-\frac{2}{3}v^3$, what corresponds to the bottom of the
potential pit, then the phase curve shrinks to a point. In the case
$-\frac{2}{3}v^3<H<0$ we have closed phase curves representing
cnoidal waves of the form:
\begin{equation}\label{EQ7}
x^{m}(t)=a+b\cdot cn^2(kt\mid m),
\end{equation}
where $cn(z\mid m)$ denotes the Jacobi elliptic cosine function of modulus $m$.
Substitution of (\ref{EQ7}) in (\ref{EQ4}) gives:
\begin{equation}\label{EQ9}
\left\{
\begin{array}{l}
x^{m}(t)=v+v\left[ (1-2m)+3m\cdot cn^2(kt\mid m)\right]/{\sqrt{m^2-m+1}}
\\ T^m=4K(m)/k
\\ k=\frac{1}{2}\sqrt{v/\beta}(m^2-m+1)^{-\frac{1}{4}}
\end{array}
\right.
\end{equation}
Here $K(m)$ is the complete elliptic integral of the first kind, $T^m$ is the period of the corresponding solution, and $m$ is the modulus of elliptic functions. Under $H=0$ $(m=1)$, what corresponds to a separatrix on a phase plane, solution (\ref{EQ9}) yields the homoclinic orbit:
\begin{equation}
x(t)=\frac{3v}{\cosh ^2 \left( \frac{t}{\Delta} \right)},
\end{equation}
where $\Delta =2\sqrt{\frac{\beta }v}$ \cite{book2}.

\section{A criterion for chaotic motion}

Let's transform variables $\left( x,p\right)$ to the canonical pair action -
angle ($I$ - action, $\theta$  - angle) using standard formulae:
\begin{equation}\label{EQ10}
I=\frac 1{2\pi }\oint p(x,H)dx=I(H),\theta =\frac{\partial S(x,I)}{\partial I%
},S(x,I)=\int p(x,H)dx,
\end{equation}
where $S(x,I)$ is the reduced action. The Hamiltonian of a perturbed
motion we can write in the form:
\begin{equation}\label{EQ11}
H(I,\theta ,t)=H_0(I)+V(I,\theta ,t),
\end{equation}
where $H_0(I)$ is the Hamiltonian of the nonperturbed system from
(\ref{EQ6}), and $V(I,\theta ,t)$ corresponds to the perturbation term in equation (\ref{EQ3}).

Assuming perturbation $V$ to be periodical on time with frequency
$\nu $, and taking into consideration the fact that nonperturbed
motion is integrable, we may expand the perturbation into the following
Fourier series:
\begin{equation}\label{EQ12}
\left\{
\begin{array}{l}
V(I,\theta ,t)= \frac {1}{2} \sum\limits_{k,l}V_{kl}(I)\exp
[i(k\theta -l\nu t],
\\ V_{kl}=V_{-k,-l}^{*}.
\end{array}
\right.
\end{equation}
The perturbed equations of motion in variables $(I,\theta )$ take now the form:
\begin{equation}\label{EQ13}
\left\{
\begin{array}{l}
{\dot I=-\frac{1}{2} i \sum\limits_{k,l}kV_{kl}(I)\exp [i(k\theta 
-l\nu t]} \\
{\dot \theta =\omega (I)+\frac{1}{2} i \sum\limits_{k,l}}\frac{dV_{kl}}
{dI}\exp [i(k\theta -l\nu t)]
\end{array}
\right.
\end{equation}
where $\omega (I)=\frac{dH_0}{dI}$ is the frequency of the nonperturbed motion.
Let's consider the motion in the vicinity of one particular resonance:
\begin{equation}
m\omega(I_{mn})-n\nu=0.
\end{equation}
The width of this resonance on frequency is $\Omega=\sqrt{4 V_0| \omega ^{^{\prime }}|}$, where $V_0\equiv |V_{mn}(I_{mn})$ and  $\omega ^{^{\prime }}\equiv d\omega(I_0)/dI$.
The distance between adjacent resonances is equal to
\begin{equation}\label{EQ17}
\Delta \omega _{\alpha \beta }=\left| \omega (I_{m\pm \alpha ,n\pm \beta
})-\omega (I_{m,n})\right|,
\end{equation}
where $\alpha =0,1$; $\beta =0,1$.
As a condition of chaotic motion, which means an appearance of
mixing, we'll use the Chirikov criterion for the overlap
of resonances \cite{book3}:
\begin{equation}\label{EQ30}
K=\left( \frac \Omega {\Delta \omega }\right) ^2\geq 1.
\end{equation}
where $\Delta \omega $ is the minimal possible value of $\Delta \omega_{\alpha\beta}$ from (\ref{EQ17}). For the cases $(\alpha,\beta)=(1,0)$, $(0,1)$ and $(1,1)$ calculations similar to \cite{book3} give respectively the following conditions for the border of stochasticity:
\begin{equation}\label{EQ31}
\left\{
\begin{array}{l}
K_{10}=4 V_0| \omega ^{^{\prime }}| m^2 / \omega ^2
\geq 1,\mbox{ }m\gg 1,n\geq 1\\
K_{01}=4 V_0| \omega
^{^{\prime }}| n^2 / \omega ^2 \geq 1,\mbox{ }n\gg 1,m\geq 1\\
K_{11}=4 V_0| \omega ^{^{\prime }}| m^2n^2 / \omega
^2(m-n)^2 \geq 1,\mbox{ }m,n\gg 1,\mbox{ }\left| m-{}n\right| \sim
1
\end{array}
\right.
\end{equation}

\section{Chaos of Korteweg-de Vries waves}

To apply these results to the system, governed by KdV equation, 
one must first evaluate the
exact expressions for $\omega $ and $\omega ^{^{\prime }}$, which
appear in the conditions (\ref{EQ31}). The frequency $\omega $ is equal to
\begin{equation}
\omega =\frac{2\pi }T=\frac{\pi k}{2K(m)}.
\end{equation}
For convenience let's rewrite it as:
\begin{equation}\label{EQ36}
\omega =\frac{\pi}{K(z)}\sqrt{\frac{c \sin{\frac{\phi}{3}}}{2\beta\sqrt{3}}},
\end{equation}
where we introduced $\varphi$ by the correlation $\cos{\frac{\varphi}{3}}=1+3\frac{H}{c^3}$, and $z\equiv \frac{1}{2}-\frac{\sqrt{3}}{2}\cot{\frac{\varphi}{3}}$.
To evaluate $\omega ^{^{\prime }}$, one should represent it as
\begin{equation}
\omega ^{^{\prime }}=\frac{d\omega }{dH}\frac{dH}{dI}=\omega \frac{d\omega}{d\varphi}\left( \frac{dH}{d\varphi}\right) ^{-1}.
\end{equation}
Tedious calculations give finally
\[
\omega ^{^{\prime }}=\frac{\pi ^2\omega}{16c^{5/2}
\sin \varphi K^2(z)}\sqrt{\frac{3}{2\beta \sin \frac \varphi 3}}\left[ -4\cos \frac \varphi 3 F\left( 1.5;0.5;1;z \right)
+\right.
\]
\begin{equation}\label{EQ37}
\left. +\left( \sqrt{3}\sin \frac \varphi 3+\cos \frac \varphi
3\right)F\left( 1.5;1.5;2;z\right) \right].
\end{equation}
Substituting (\ref{EQ36}) and (\ref{EQ37}) in (\ref{EQ31}) one makes sure that with approach to
separatrix $K\rightarrow $ $\infty $ under arbitrary small
$V_0 $. This means that for any external perturbation,
however small it is, Chirikov criterion is executed starting with
some energies, and this leads to the formation of a corresponding
stochastic layer. The transition from regular to chaotic motioncan be found
approximately from the condition $K\approx 1$. This border is defined on energy by the inequality:
\begin{equation}\label{EQ38}
H_{\min }\leq H\leq 0,H_{\min }=\frac{c^3}3(\cos \varphi _{\min
}-1).
\end{equation}
Here the value of $\varphi _{\min }$ is found from:
\begin{equation}
\frac{4v^3\sin {\varphi_{\min }} \tan (\varphi_{\min
}/3)K(z_{\min })}{-4{}F\left(1.5; 0.5;1;z_{\min }\right) +\left(
\sqrt{3}\tan ( \varphi_{\min }/3) +1\right) {}F\left(
1.5;1.5;2;z_{\min }\right) }=\pi V_0\zeta\left(m,n\right),
\end{equation}
in which $z_{\min }=\frac 12-\frac{\sqrt{3}}2\cot
\frac{\varphi_{\min }}{3}$ and $\zeta(m,n)=m^2$, $n^2$,
$\frac{m^2n^2}{(m-n)^2}$ for three cases in (\ref{EQ31}).
Certainly, there is only half-width of stochastic layer, but the phase curves, laying out separatrix don't constitute an interest, being physically  meaningless
(they are unbounded on infinity).

\section{Summary and conclusions}

We have calculated the width of the stochastic layer around a separatrix, corresponding to a soliton solution of Korteweg-de Vries equation under hamiltonian perturbations. As far as the motion in this stochastic layer is chaotic ( in the meaning of mixing), so nonlinear wave solutions, which correspond to the phase curves within this stochastic layer will also posess certain chaotic properties. Let's now consider the question about the spatial region, where this chaotic behaviour can be registered.

The time length, after which stochasticity in a nonlinear oscillator can be found is defined by $\tau _c\ll t$, where $\tau _c$ is a time of the decay of correlations. In the work \cite{book3} the following estimation of
$\tau _c$ for a nonlinear oscillator is obtained:
\begin{equation}
\tau _c\sim \frac 1{\ln K},
\end{equation}
where $K$ is a coefficient for the overlap of resonances, obtained in the previous section. As far as in our consideration time $t$ for a
nonlinear oscillator corresponds to the expression ($x-ct$) for
nonlinear waves, so the region, where chaotic regimes for these waves can be detected, is:
\begin{equation}
\frac 1{\ln K}\ll \left( x-ct\right).
\end{equation}
Therefore, we can conclude that this region represents wave formation, which
outstrips nonlinear wave or soliton, and propagates in the same direction 
with the same velocity. Under $t=0$ we obtain:
\begin{equation}
\frac 1{\ln K}\ll x .
\end{equation}
As far as with approach to a separatrix $K\rightarrow $ $\infty $,
so for solitons the minimal distance on which chaotic behaviour should appear $x _c\rightarrow 0$, and this effect
will be realized in a close vicinity of a soliton peak. The
chaos, which we are searching for, will be manifested in
the irregular small deflection from a smooth initial soliton profile.
So, as a main result of this work, it can be inferred that long-period nonlinear waves and especially solitons of Korteweg-de Vries equation obtain chaotic properties in the presence of periodic hamiltonian external perturbations.

Exactly this result was observed in an experiment with ion-acoustic
and Langmuir waves in a nonmagnetized plasma \cite{in-proc2}. There was
registered a faint splash directly before the soliton. From the
concepts developed in this paper this phenomenon can be explained
in the following way. Together with soliton wave generator also produces a group of waves of small amplitude and almost zero frequencies. These waves are produced constantly with generator
working, and therefore can be considered as a small deterministic
periodic external perturbation. Stochastic destruction of soliton
due to the influence of these waves, as it was described above,
must realize itself just before the soliton peak, exactly what was
registered in the experiment.

\section* {\bf Acknowledgements}

The author would like to thank Prof. M. Bestehorn for helpful discussions. He also acknowledges the referee for suggestions, which helped to make the paper clearer.


\begin{thebibliography}{20}
\bibitem{paper1} R .Grimshaw, X. Tian: {\em Proc.\ R. \ Soc. \ London A} {\bf 445}, 1-21 (1994).
\bibitem{paper2} Q. Cao, K. Djideli, W.G. Price, E.H. Twizell: {\em Physica D} {\bf 125}, 201-221 (1999).
\bibitem{paper3} X. Tian, R. Grimshaw: {\em Int.\ J.\ Bif.\ Chaos} {\bf 5}, 1221-1233 (1995).
\bibitem{paper4} R. Grimshaw, X. Tian: {\em Physica D} {\bf 77}, 405-433 (1994).
\bibitem{in-proc1} T. Kawahara, S. Toh: {\em On some properties of solutions to a nonlinear evolution equation including long-wavelength instability},
in Nonlinear Wave Motion~43, ed. A. Jeffrey, Longman, New York 1989.
\bibitem{paper5} T. Ogawa: {\em Hiroshima \ Math.\ J.} {\bf 24}, 401-422 (1994).
\bibitem{book1}  A. J. Lichtenberg, M. A. Liberman: {\em Regular and
Chaotic Dynamics}, 2nd ed., Springer-Verlag, New York 1992.
\bibitem{in-proc2}  H. Ikezi: {\em Experiments on Solitons in Plasmas}, in Solitons in
Action, K. Longren and A.C. Scott eds., Academic Press, New York 1978.
\bibitem{book2} V. I. Karpman: {\em Nonlinear waves in dispersive media}, Pergamon Press, New York 1973.
\bibitem{book3} R. Z. Sagdeev, D. A. Usikov, G. M. Zaslavsky: {\em Nonlinear physics: From the pendulum to turbulence and chaos}, Harwood Academic Press, New York 1988.
\end{thebibliography}
\end{document}